\documentclass[fleqn,twoside]{article}
\usepackage{espcrc2}

\usepackage{graphicx}
\usepackage[figuresright]{rotating}

\newcommand{\AmS}{{\protect\the\textfont2
  A\kern-.1667em\lower.5ex\hbox{M}\kern-.125emS}}

\title{On Supersymmetries}

\author{Chun Liu\address[ITP]{Institute of Theoretical Physics, Chinese 
Academy of Sciences, P. O. Box 2735, Beijing, China}%
        \thanks{liuc@itp.ac.cn}}
       
\begin{document}

\begin{abstract}
After reviewing electroweak (EW) scale supersymmetry (susy) and split 
susy, as well as their implications in very high energy cosmic rays, 
I present a high scale susy model for fermion masses.  An 
$\mathcal O(0.1)$ $\nu_e-\nu_{\tau}$ mixing is expected. 
\vspace{1pc}
\end{abstract}

\maketitle

\section{Overview of susy}

\subsection{EW scale susy}

Originally, EW scale susy was introduced because of grand unification 
and naturalness.  LEP may imply grand unification of the three gauge 
couplings at $M_{GUT}\sim 10^{15-16}$ GeV.  One generation fermions 
compose of the $SO(10)$ ${\bf 16}$ representation with neutrino masses 
$m_{\nu}\sim\displaystyle\frac{m_t^2}{M_{GUT}}$.  The Standard Model 
(SM) Higgs mass $m_h \ll M_{GUT}$, it is unnatural from quantum 
mechanics!  EW scale susy is very required. 
                                                                                
Consider very high energy cosmic rays.  After hitting the atmosphere they 
can produce $100$ GeV susy particles, $E_{CM}=\sqrt{2m_pE_{CR}}$ can be 
as large as $100$ TeV.  However, this is LHC-like with a high background.  
Specific models may have distinct signals in ultra high energy cosmic 
rays.  
                                                                                
One of the guiding problems in susy studies is the flavor changing 
neutral current (FCNC) problem.  For an example, $\mu\to e\gamma$ 
requires certain pattern of slepton masses which result from the susy 
breaking mechanism.  

One solution to the FCNC problem is gauge mediated susy breaking 
\cite{gmsb}.  Susy breaking occurs in a hidden sector, the breaking 
transfers to the SM sector via gauge interactions, sleptons are 
degenerate.  Taking the susy breaking scale as $\sqrt{F}$, 
$m_{slepton} \sim\displaystyle \frac{\alpha}{4\pi}\frac{F}{M}$ with 
$M$ being messenger masses, $\sqrt{F}\geq 100$ TeV.  The gravitino mass 
is $m_{gravitino} \sim\displaystyle \frac{F}{M_{planck}} \ll $ EW scale.  
Such a gravitino can be the dark matter!  The next lightest stable susy 
particle (NLSP) is stau: 
$\tilde{\tau}_R$ ($\sim 100$ GeV).  Its lifetime 
($\tilde{\tau}\to\tau+$ gravitino) 
$\tau_{\tilde{\tau}_R} \sim\displaystyle \frac{16\pi F^2}{m_{\tilde{\tau}_R}^5}$ 
which can be long enough ($1$ sec)!  

Long-lived charged particles can be seen in cosmic rays.  In Ref. 
\cite{abc}, by taking 
$\sqrt{F}\simeq 5\times 10^6-5\times 10^8~{\rm GeV}$, it was obtained 
that 
\begin{equation}
\label{1}
c\tau_{\tilde{\tau}_R} \sim \left(\frac{\sqrt{F}}{10^7~{\rm GeV}}
\right)^4\left(\frac{100 {\rm GeV}}{m_{\tilde{\tau}_R}}\right)^5 10\,
{\rm km}\,.   
\end{equation}
The stau production cross section is smaller than SM processes (muon 
production) by about 2 orders of magnitude.  But, the long lifetime 
compensates for the small production.  Neutrino telescopes can detect 
these charged NLSPs.  The effective detect range is hundred or thousand 
kilometers in IceCube.  Staus are pair produced.  Typical signals are 
two tracks separated by about $100$ m.  The di-muon background is not 
significant.  A few events are expected per year in IceCube.  A similar 
analysis of a model with a quintessino LSP is presented in the Symposium 
\cite{zm}.  

\subsection{Split susy}
                                                                                
However, there maybe no naturalness.  The cosmological constant with 
$10^{120}$ fine tuning might be just so from the anthropic point of 
view.  Then SM is just the full theory.  It is a pity for theorists if 
there is no susy.  Split susy was proposed to give up naturalness while 
keeping good points of low energy susy.  

Split susy keeps all fermions to be around the EW scale, and takes 
sfermions heavy $\geq 100$ TeV {\bf except for one light Higgs}.  Then 
it has GUT and the dark matter!  It trivially has no any FCNC problem.  

The gluino is very long-lived because it decays via virtual squarks.  
Its lifetime ($\tilde{g}\to$ quark + anti-quark + LSP) 
\begin{equation}
\label{2}
\tau_{\tilde{g}} \sim 3\times 10^{-2}~{\rm sec}
\left(\frac{m_{squark}}{10^9~{\rm GeV}}\right)^4
\left(\frac{1 {\rm TeV}}{m_{\tilde{g}}}\right)^5\,. 
\end{equation}
In Ref. \cite{hlmr}, gluino pair production in ultra high energy cosmic
rays is analyzed.  Neutrino telescopes can detect down-going pairs.  One 
event is expected in IceCube per year when $m_{\tilde{g}}< 170$ GeV.  
                                                            
In Ref. \cite{agn}, air showers caused by cosmic gluino-contained 
hadrons are analyzed.  The very low inelasticity of $\tilde{g}$-air 
interaction guarantees the primary particle retains most of its energy 
traveling to the ground, while the cascade particles behave as in 
ordinary air showers.  Pierre Auger Observatory can detect such showers 
if cosmic sources are able to accelerate particles above 
$5\times 10^{13}$ GeV.

\section{Susy for fermion masses}

If susy is not for stabilizing the EW energy scale, what else is this 
beautiful mathematical physics used for in reality?  We propose susy 
for flavors \cite{l1}!  The flavor puzzle lies in the pattern of fermion 
masses, mixing and CP violation.  It is observed that for masses of 
fermion generations: $3rd~ \gg 2nd~ \gg 1st$.  We propose a family 
symmetry: $Z_{3L}$ of the $SU(2)_L$ doublets of leptons and quarks, 
under which $L_1, Q_1 \to L_2, Q_2 \to L_3, Q_3 \to L_1, Q_1$.  It 
results in $m_{\tau} \neq 0$, $m_t \neq 0$, $m_b \neq 0$ only.  The 
crucial question is that how does $Z_{3L}$ break?  For leptons, we have 
noted \cite{dl} that {\bf sneutrino vacuum expectation value (vev) 
$v_i \neq 0$, $LLE^c$ contributes to charged lepton masses} where $E^c$ 
stands for $SU(2)_L$ singlet leptons.  But, one neutrino mass 
$m_{\nu}\sim \displaystyle \frac{(g_2v_i)^2}{M_{\tilde{Z}}}\sim 100$ MeV 
with $g_2$ being the gauge coupling and gaugino masses $M_{\tilde{Z}}$ 
around the EW scale, it is  too large!  We will make gauginos enough 
heavy.  

Let us describe the model.  In addition to gauge symmetries and susy, 
$Z_{3L}$ is assumed.  It breaks in soft terms.  In writing the 
Lagrangian, both kinetic terms and the superpotential consist of 
most general $Z_{3L}$ terms.  The canonical kinetic form is got via 
field redefinition.  Finally, the superpotential becomes
\begin{equation}
\label{3}
\begin{array}{lll}
{\mathcal W} &=& -y_{\tau} H_d L_{\tau} E^c_{\tau}
+L_eL_{\mu}(\lambda_{\tau} E^c_{\tau}+\lambda_{\mu} E^c_{\mu}) \\
   & & +\bar{\mu} H_uH_d \,,
\end{array}
\end{equation}
where $H_{u,d}$ are the two Higgs doublets, $\bar{\mu}$ a mass parameter 
and $y_{\tau}, \lambda_{\tau,\mu}$ couplings.  We see that $H_d$ 
contributes to the tau mass only; sneutrinos in $L_e$ and $L_{\mu}$ to 
the muon mass, and the electron remains massless.  

It is important to note that masslessness of the electron is kept by 
susy.  Generally, family symmetries keep the muon and electron massless.  
Once the family symmetry is broken, however, both muon and electron get 
their masses.  And there is no reason to expect a hierarchy between the 
muon mass and the electron mass.  In this model, it is the simplicity of 
the superpotential that makes the electron massless even if sneutrino 
vevs are non-vanishing.  The simplicity comes from susy.  The 
non-vanishing electron mass is therefore due to susy breaking effects.  

Soft susy terms breaking $Z_{3L}$ can be written.  It has been shown 
that by fine-tuning, one Higgs scalar doublet with a mass-squared 
$-m_{EW}^2$ can be obtained.  EW symmetry breaking is achieved.  The 
tuning is at the order of $m_S^2/m_{EW}^2$ where $m_S$ is the susy 
breaking scale.  This light Higgs is a mixture of $H_u$, $H_d$ and 
sleptons in Eq. (\ref{3}).  In terms of the latter fields, 
\begin{equation}
\label{4}
v_u \neq 0\,, ~v_d\neq 0\,, ~ v_{l_{\alpha}} \neq 0 
~(\alpha = e,~\mu,~ \tau)\,.
\end{equation}
Numerically $v_d\sim 10$ GeV and $v_{l_{\alpha}}\sim 1$ GeV.  

Is a large $v_{l_{\alpha}}$ safe?  In addition, huge slepton-Higggs 
scalar mixing mass-squared soft terms induce large lepton-Higgsino 
mixing at the loop-level, 
$m_{\alpha h}=\displaystyle\frac{g_2^2B_{\mu_{\alpha}}}{16\pi^2M_{\tilde{Z}}}$
which is about $10^{-3}m_S$.  The neutralino  
$(\nu_e~\nu_{\mu}~\nu_{\tau}~\tilde{h}_d^0~\tilde{h}_u^0~\tilde{Z})$ 
mass matrix is 
\begin{equation}
\label{5}
\left(
\begin{array}{cccccc}
0       & 0          &0            &0         & m_{eh}   &av_{l_e}    \\
0       & 0          &0            &0         &m_{\mu h} &av_{l_{\mu}}\\
0       & 0          &0            &0         &m_{\tau h}&av_{l_{\tau}}\\
0       & 0          &0            &0         &-\bar{\mu}&av_d \\
m_{eh}  &m_{\mu h}   &m_{\tau h}   &-\bar{\mu}&0         &-av_u\\
av_{l_e}&av_{l_{\mu}}&av_{l_{\tau}}& av_d     &-av_u     &M_{\tilde{Z}}
\end{array}
\right)
\end{equation}
where $a=\displaystyle(\frac{g_2^2+g_1^2}{2})^{1/2}$, and $\tilde{h}$ 
stands for higgsinos.  Its large eigenvalues are the following 
\begin{equation}
\label{6}
\Lambda_1 \simeq M_{\tilde{Z}} \,,~\Lambda_2 \simeq \bar{\mu} \,,~
\Lambda_3 \simeq -\bar{\mu} \,.
\end{equation}
There are three light neutrinos.  This is a realization of the see-saw 
mechanism with heavy higgsinos and gauginos playing the role of 
right-handed neutrinos.  The light Majorana neutrino mass matrix is then 
\begin{equation}
\label{7}
\begin{array}{lll}
m^{\nu}&\simeq&-m_{\rm Dirac} M_R^{-1} m_{\rm Dirac}^T \,, \\
[3mm]
& = & \displaystyle - \frac{a^2}{M_{\tilde{Z}}}
\left(
\begin{array}{ccc}
v_{l_e}v_{l_e}      &v_{l_e}v_{l_{\mu}}      &v_{l_e}v_{l_{\tau}}     \\
v_{l_{\mu}}v_{l_e}  &v_{l_{\mu}}v_{l_{\mu}}  &v_{l_{\mu}}v_{l_{\tau}} \\
v_{l_{\tau}}v_{l_e} &v_{l_{\tau}}v_{l_{\mu}} &v_{l_{\tau}}v_{l_{\tau}}
\end{array}
\right) \,.
\end{array}
\end{equation}
It is a democratic matrix for neutrinos, naturally large neutrino mixing 
are expected.  The nonvanishing mass is
$m_{\nu_3}=\displaystyle\frac{a^2}{M_{\tilde{Z}}}
v_{l_{\alpha}}v_{l_{\alpha}}\sim 10^{-1}-10^{-2}$
eV when $M_{\tilde{Z}}\sim 10^{11}-10^{12}$ GeV. 

The electron mass is due to soft susy breaking terms, 
$\delta M^l_{\alpha\beta}\simeq\displaystyle 
\frac{\alpha}{\pi}\frac{y_{\tau}\tilde{m}_S v_d}{m_S}$, 
with $\tilde{m}_S$ being trilinear soft masses.  Taking 
$\tilde{m}_S/m_S\simeq 0.1$, 
$\delta M^l_{\alpha\beta} \sim {\cal O}$(MeV).

{\bf Neutrino oscillation} should be analyzed.  
One SM singlet superfield $N$ is needed, 
\begin{equation}
\label{8}
{\mathcal W} \supset \kappa_{\tau} H_uL_{\tau}\bar{N}
+ \tilde{M}\bar{N}\bar{N}+\kappa_d H_uH_d\bar{N}
+\tilde{\kappa}_3 \bar{N}^3 \,,
\end{equation}
where $\kappa$'s are couplings and $\tilde{M}$ the mass of $N$.  The 
full neutrino mass matrix is
\begin{equation}
\label{9}
\begin{array}{lll}
{\mathcal M}^{\nu} & = & \displaystyle - \frac{a^2}{M_{\tilde{Z}}}
\left(
\begin{array}{ccc}
v_{l_e}v_{l_e}      &v_{l_e}v_{l_{\mu}}      &v_{l_e}v_{l_{\tau}}     \\
v_{l_{\mu}}v_{l_e}  &v_{l_{\mu}}v_{l_{\mu}}  &v_{l_{\mu}}v_{l_{\tau}} \\
v_{l_{\tau}}v_{l_e} &v_{l_{\tau}}v_{l_{\mu}} &v_{l_{\tau}}v_{l_{\tau}}+x
\end{array}
\right)
\end{array}
\end{equation}
with $x$ being 
$\displaystyle\frac{M_{\tilde{Z}}}{\tilde{M}}\left(\frac{\kappa_{\tau}v_u}{a}\right)^2$.
Its eigen values are
\begin{equation}
\label{10}
\begin{array}{lll}
m_{\nu_3}  & \simeq & \displaystyle \frac{a^2}{M_{\tilde{Z}}}
v_{l_{\tau}}^2 + \frac{(\kappa_{\tau}v_u)^2}{\tilde{M}} \,, \\
m_{\nu_2}  & \simeq & \displaystyle \frac{a^2}{M_{\tilde{Z}}}
(v_{l_e}^2+v_{l_{\mu}}^2)\frac{x}{x+v_{l_{\tau}}^2} \,, \\
m_{\nu_1}  & =      & 0                        \,.
\end{array}
\end{equation}
Solar neutrino problem requires that $m_{\nu_2}\simeq (10^{-2}-10^{-3})$ 
eV which is achieved when $M_{\tilde{Z}}\sim 10^{13}$ GeV.
Atmospheric neutrino problem requires a certain cancellation between
the terms $\displaystyle\frac{a^2}{M_{\tilde{Z}}}v_{l_{\tau}}^2$ and
$\displaystyle\frac{(\kappa_{\tau}v_u)^2}{\tilde{M}}$, in order to make
$m_{\nu_3}\sim 10 ^{-1}-10^{-2}$.  The lepton mixing are 
\begin{equation}
\label{11}
|V_{e2}| = \frac{v_{l_{\mu}}^2-v_{l_e}^2}{v_{l_e}^2+v_{l_{\mu}}^2}
\simeq O(1)  \,.
\end{equation}
\begin{equation}
\label{12}
|V_{\mu3}| \simeq \frac{|\lambda_{\tau}|\sqrt{v_{l_e}^2+v_{l_{\mu}}^2}}
{\sqrt{y_{\tau}^2 v_d^2+|\lambda_{\tau}|^2(v_{l_e}^2+v_{l_{\mu}}^2)}}\,.
\end{equation}
The maximal mixing can be achieved.  The $\nu_e-\nu_{\tau}$ mixing is 
\begin{equation}
\label{13}
|V_{e3}| \simeq \frac{v_{l_{\mu}}^2-v_{l_e}^2}
{\sqrt{v_{l_e}^2+v_{l_{\mu}}^2}v_{\tau}} \,.
\end{equation}
It is $\sim 0.1$ if $\sqrt{v_{l_e}^2+v_{l_{\mu}}^2}/v_{\tau} \sim 0.1$.
                                                                                
{\bf Quark masses} also have three origins: 
{\it Higgs vevs, sneutrino vevs, soft trilinear $Z_{3L}$ violating terms.}  
However, the role of the sneutrino vevs and soft trilinear terms are 
switched.  Sneutrino vevs contribute to the first generation quark mass, 
and soft trilinear $Z_{3L}$ violating terms to charm and strange quark 
masses.  The hierarchy between the second and first generation is not 
automatic.  A special structure of soft breaking terms of squarks 
is needed.  A good point is that $m_u < m_d$ can be understood.

{\bf Higgs} mass $\simeq 145\pm7$ GeV.  This was obtained in Ref. 
\cite{li}, which considered a high scale susy scenario in a different 
physics content.  One specific point is that we now have 
$\tan\beta\simeq m_t/m_b$.  

Coming back to cosmic ray physics, our model has little to say compared 
to SM.  One special point is that we predict a relatively large 
$\theta_{13}$.  This has certain cosmic ray physics implication 
\cite{xing}.  \\

The author acknowledges the support of the National Natural Science 
Foundation of China.

\end{document}